# 3D Fluorescent Mapping of Invisible Molecular Damage after Cavitation in Hydrogen Exposed Elastomers.

Xavier P. Morelle*,†,‡,1, Gabriel E. Sanoja†,1, Sylvie Castagnet2, and Costantino Creton*,1,3.

[1] SIMM, ESPCI Paris, Université PSL, CNRS, Sorbonne Université, 75005, Paris, France.

[2] Institut Pprime (UPR 3346 CNRS – ENSMA – Université de Poitiers), Department of Physics and Mechanics of Materials, 1 Avenue Clément Ader, BP 40109, 86961 Futuroscope cedex, France

[3] Global Station for Soft Matter, Global Institution for Collaborative Research and Education, Hokkaido University, Sapporo, Japan.

**ABSTRACT:** Elastomers saturated with gas at high pressure suffer from cavity nucleation, inflation, and deflation upon rapid or explosive decompression. Although this process often results in undetectable changes in appearance, it causes internal damage, hampers functionality (*e.g.*, permeability), and shortens lifetime. Here, we tag a model poly(ethyl acrylate) elastomer with π-extended anthracene-maleimide adducts that fluoresce upon polymer chain scission, and map in 3D the internal damage present after a cycle of gas saturation and rapid decompression. Interestingly, we observe that each cavity observable during the decompression results in a damaged region, the shape of which reveals a fracture locus of randomly oriented penny-shape cracks (*i.e.*, with a flower-like morphology) that contain crack arrest lines. Thus, cavity growth likely proceeds discontinuously (*i.e.*, non-steadily) through the stable and unstable fracture of numerous 2D crack planes. This non-destructive methodology to visualize in 3D molecular damage in polymer networks is novel and serves to understand how fracture occurs under complex 3D loads, predict mechanical aging of pristine looking elastomers, and holds potential to optimize cavitation-resistant materials.

## Introduction

Elastomers are ubiquitous in applications going from classical rubber tyres and seals, to more recent ones like soft robotics [1-3], and wearable electronics [4, 5]. Among classical applications, one of the most technologically relevant is that of seals where the ability of elastomers to undergo large reversible deformations is used to effectively block gas or fluid transport (*i.e.*, the classic O-ring gasket). Gas poses specific problems since it can diffuse, sorb, and saturate the elastomer at high pressure. Upon a sudden drop in pressure, the resulting imbalance in chemical potentials drives gas diffusion at a rate that is generally too slow to prevent solid-gas phase separation, leading to cavity nucleation and growth. This process, known as *rapid* or *explosive decompression*, has been observed in elastomers saturated with gases [6] and more recently with liquids [7], and is of recent concern for the storage and transportation of compressed hydrogen fuel because it hampers performance (*e.g.*, permeability) and ultimately leads to failure [8].

Cavity nucleation and growth has been previously investigated in elastomers [9, 10], and more recently in hydrogels [11-13]. Although this was first perceived as reversible deformation of a preexisting crack or flaw [9, 14, 15], it was later ascribed to irreversible fracture [16-22] by polymer network chain scission [7, 22]. This interpretation is clear when large cavities result in surface blisters of size ~ 100-1000 μm [9, 19, 23], but is less obvious when small cavities completely disappear upon unloading. The more recent theoretical models have based their cavitation criterion not only on elastic considerations but also on the strength [24], strain hardening and fracture toughness of elastomers [17, 18]. Some models have assumed that fracture by cavitation results from diffuse (*i.e.* radial) scission of polymer chains along the cavity wall [16, 17, 25], whereas others have argued that it follows from the localized growth of a penny-shaped crack [18, 21, 24]. However, the *inability to directly visualize the damage* to the network by chain scission still raises question about the precise shape of the damage zone that results from cavity nucleation and growth [13]. This information could be used to mitigate overengineering, predict lifetime, and prevent mechanical failure during use.

Detecting and visualizing damage by network chain scission in elastomers has historically been challenging because elastic restoring forces and surface tension close cracks and flaws in the absence of a load (*i.e.*, the unloaded state). If recent *in-situ* micro-tomography studies have enabled 3D visualization of cavities opened by hydrostatic tension through mechanical loading of a poker-chip geometry [26], or with a rapid gas decompression set-up [27], techniques based on X-rays transmission cannot be used to detect closed cavities, whereas others based on emission of acoustic waves lack 3D spatial resolution and have low signal detection thresholds [28]. Polymer mechanochemistry has recently proven promising to optically detect and map forces on molecules and directly detect chemical bond scission [29-34]. Slootman et al. [35] have recently shown that mechanofluorescent crosslinkers can be used to spatially map and quantify the concentration of broken bonds in elastomers with a few microns spatial resolution, and used it to demonstrate the rate- and temperature-dependence of the damage zone size associated with a propagating crack. Applied to cavitation, such a method would be particularly useful to visualize damage after full unloading when cavities become invisible with conventional optical techniques.

Here, we used a similar methodology and tagged elastomers with mechanofluorescent crosslinkers (*i.e.*, mechanophores) based on π-extended anthracene-maleimide adducts. Upon polymer chain elongation until failure, these crosslinkers undergo a [4+2] cycloreversion reaction that results in fluorophores of high quantum yield and stability to photobleaching (**FIG1**A). Interestingly, we found in fracture experiments that, although this cycloreversion reaction has a slightly lower activation energy than the dissociation of a C-C bond, scission of the mechanofluorescent crosslinker quantitatively reports for network failure by chain scission [35, 36]. As such these probes are suitable for *mapping molecular damage in the unloaded state after rapid decompression and gas desorption*.

We synthesized a model tagged elastomer, $EA_{0.5}(+)$, by free radical polymerization in the bulk (no added solvent) of ethyl acrylate monomer, 1,4-butanediol diacrylate and mechanophore crosslinkers, and 2-hydroxy-2-methylpropiophenone initiator (for synthetic details refer to Materials & Methods). The resulting material has properties like those of conventional elastomers with a glass transition temperature $T_g \approx -18°C$, Young's modulus $E \approx 1$ *MPa*, fracture toughness $G_c \approx 180$ $J/m^2$, and stress-strain curves under uniaxial and pure shear deformation like those of its untagged counterpart, $EA_{0.5}(-)$ (see **FIG S1** in the SI).

Then, we first saturated $EA_{0.5}(+)$ with $H_2$ gas at a pressure of 5 MPa for 10 hours and then decompressed it with a set-up similar to that of Gent and Tompkins [6] while monitoring cavity nucleation and growth with an optical camera [37] (orange regime of the pressure cycle in **FIG1**B). At the end of the cycle, the resulting specimen, now pristine looking again and equilibrated at atmospheric pressure, was imaged with a laser scanning fluorescent microscope ($\lambda_{ex} = 405$ *nm*, green regime in **FIG1**B) to reconstruct 3D maps of cavitation-induced damage zones.

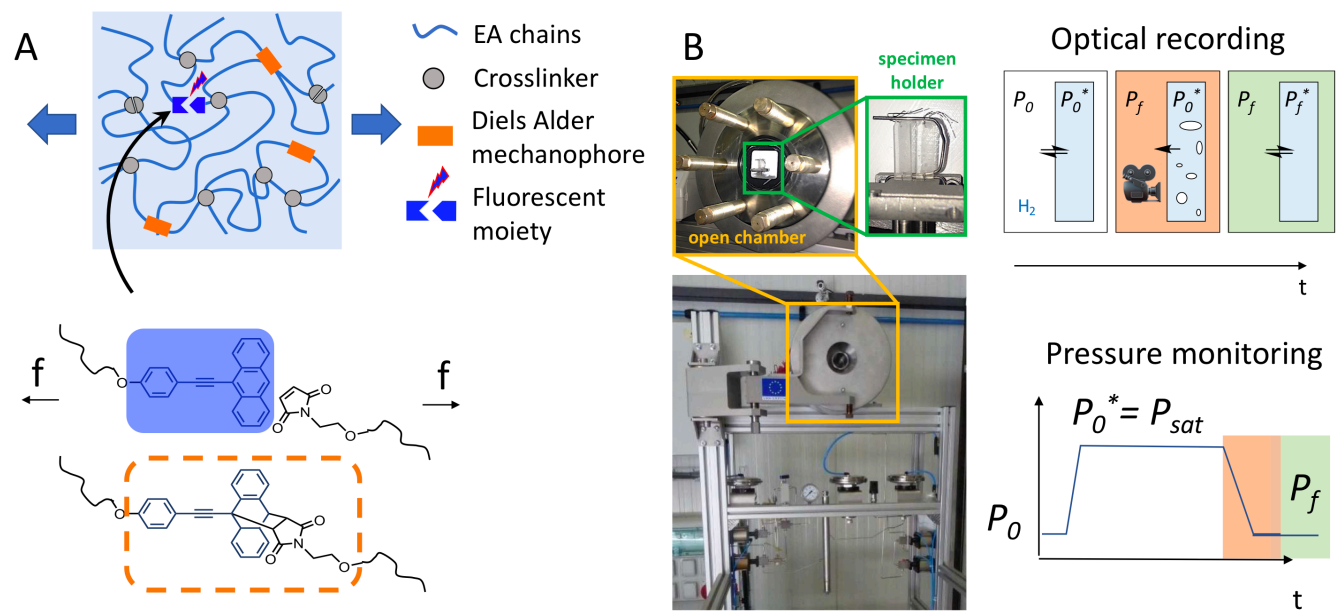

**Figure 1.** (A) *Incorporation of mechanophores in a poly(ethyl acrylate) elastomer, $EA_{0.5}(+)$. Upon polymer chain scission, the π-extended anthracene-maleimide probes undergo a cycloreversion reaction that results in fluorescent moieties, reporting bond scission representative of the network failure. (B) Rapid decompression set-up: test was performed in a pressure chamber [38] where an $EA_{0.5}(+)$ specimen was saturated with hydrogen at a pressure $P_{sat} = 5$ MPa for 10 h. Upon rapid decompression, the hydrogen undergoes solid-gas phase separation leading to the nucleation of cavities that first grow and then progressively disappear over time. Cavity growth was monitored in the orange regime, whereas damage was mapped in the green regime.*

## Cavity Nucleation & Growth by Rapid Decompression

**FIG2**A displays optical snapshots of $EA_{0.5}(+)$ subjected to rapid decompression. The specimen was saturated with hydrogen and then decompressed to 0.1 $MPa$ at a rate of 10 $MPa/min$. After ~ 60 $s$ of decompression, cavities appeared throughout the specimen, progressively grew, reached a maximum size (red squares in **FIG2**B), and disappeared over time. Most of these cavities first appeared circular and then transitioned to an anisotropic shape (blue dots in **FIG2**B), but some small ones remained circular over the entire course of decompression. At the maximum size, the average effective radius, r = √(cavity surface area/π), was 163 $\mu m \pm 6$ $\mu m$ (see distribution in **FIG2**C, and image processing details in Materials & Methods), with smaller cavities disappearing faster than large ones. This time of cavity disappearance is controlled by the rate of hydrogen diffusion and depends on the distance between the cavity nucleus and the specimen surface. However, this distance, as the 3D orientation of cavities relative to the observation plane, is not measurable with the 2D images of **FIG2**A. Visualizing damage with a laser scanning confocal microscope provides additional 3D spatial information on the location and orientation of the damage zone created by the cavity during growth.

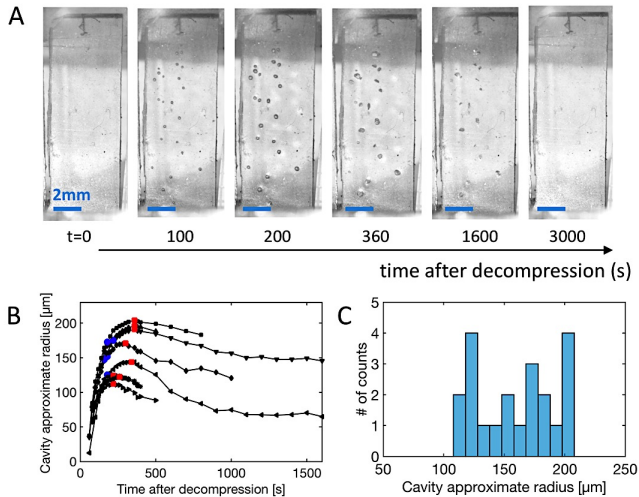

**Figure 2.** (A) ) *Optical snapshots of cavity nucleation and growth after rapid decompression at 10* MPa/min *of an $EA_{0.5}(+)$ elastomer saturated with hydrogen at 5* MPa *for 10* h *(see **Movie S1** in SI). (B) Evolution of the cavity radius (r=√(cavity surface/π)) after rapid decompression. Red squares correspond to the maximum cavity size, whereas blue dots indicate the transition from circular to anisotropic growth shape. (C) Distribution of cavity radii at the maximum size.*

## Visualization of Damage by Network Chain Scission

After rapid decompression and complete gas desorption, the pristine looking specimen was removed from the chamber and imaged with a laser scanning confocal microscope to build 3D fluorescent maps of damage by network chain scission. Interestingly, *only the regions that experienced cavity nucleation and growth exhibit zones of damaged-induced fluorescence* with size, shape, and orientation analogous to that of the inflated cavities (**FIG3** and **Movie S2**). This observation indicates that there is a one-to-one correspondence between the optical observation of an opened cavity and the detected damage by network chain scission, at least at length scales above ~ 3 $\mu m$, the resolution of the confocal microscope. The extent of damage can, in principle, be quantified from the fluorescence intensity [35, 36] but in the present case this is challenging because of the difficulty in accounting for depth dependent laser attenuation of the fluorescence in regions with arbitrary depth and orientation.

Closer inspection of these 3D fluorescent maps reveals that *each damaged region has a flower-like morphology* composed of a collection of randomly oriented penny-shape disks located around a nucleus of high fluorescence (see **FIG3**B and **Movie S2**). These damaged regions are detectable down to dimensions of ~ 30 $\mu m$, where the maximum cavity size is comparable to the resolution of the optical camera (**FIG S2**). In addition, they are enclosed in 3D boxes of highly planar aspect ratio ($\geq$ 5:1 with width ~ 100-400 $\mu m$, and height ~ 30-50 $\mu m$) and are consistently 70-80% smaller than the corresponding inflated cavities (**FIG S3**), in agreement with the unloaded state (i.e. elastic retraction) of the specimen after rapid decompression and gas desorption. Interestingly, the thickness of a penny-shape disk ~ 5-25 $\mu m$ is comparable to the length scale over which the same $EA_{0.5}(+)$ elastomer is damaged after crack propagation under monotonic loading [35, 36], indicating that *cavity nucleation and growth does result from fracture of numerous 2D crack planes in the bulk* (*i.e.*, localized fracture). This picture is reasonable considering the large strains experienced at a cavity wall and the inability of the elastomer to significantly dissipate energy by sacrificial bond scission prior to fracture.

Such damaged regions with flower-like morphology are also consistent with the anisotropic shape of the cavities that grow in PDMS under volume-controlled conditions [39], as well as with the ellipsoidal fracture surface that results from either loading a mechanically confined (*i.e.*, poker-chip) polyurethane [19] or explosively decompressing an EPDM rubber [40]. As such, tagging elastomers with mechanofluorescent probes serves here to non-destructively map damage by network chain scission accompanying the opening and closing of cavities in pristine looking elastomers, but more generally to illustrate typical features of fracture by cavitation in elastomers.

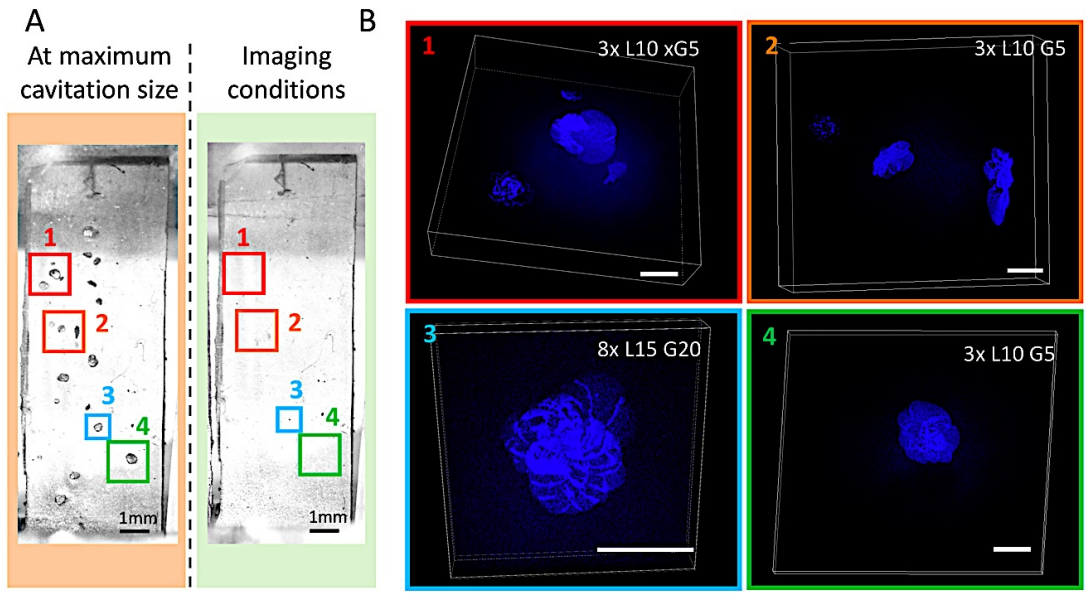

**Figure 3.** (A) *Optical snapshots of an $EA_{0.5}(+)$ elastomer at the time of maximum cavity expansion (orange) and at the time of cavity disappearance after rapid decompression and complete gas desorption (green). Clearly, the specimen is pristine looking when imaged under the laser scanning confocal microscope. (B) 3D maps of damage-induced fluorescence in regions that experienced cavity nucleation and growth. Optical magnification, laser intensity, and gain are indicated for each scan. White scale bars correspond to 200* μm.

**From Damage by Network Chain Scission to Cavity Growth Process**

Because of its good spatial resolution and more importantly This novel information on damage by network chain scission can be used to outline a mechanism of cavity growth. For that purpose, we compared optical (**FIG4**) and fluorescent images (**FIG5**) of a cavity-damaged region (n°3 in **FIG3**) that is fortuitously aligned parallel to the specimen surface, and has negligible out-of-plane tilt effects.

After rapid decompression, the competing rates of solid-gas phase separation and diffusion lead to cavity nucleation and growth under varying pressure and volume. As such, the cavity growth rate progressively decreases with cavity size (see slope of **FIG2**B) and the evolution of the cavity shape can be optically monitored with sufficient resolution only at later stages of cavity growth (*i.e. t > 100 s*, see **FIG4**A). Typically, after ~ 150-200 *s* there is a transition from circular to anisotropic shape with "popping" events that lead to concavities and a raspberry-like 3D growth (see dashed lines in **FIG4**A, and schemes *(ii)* in **FIG4**B). This breakage of symmetry has been ascribed to fracture in other elastomers and hydrogels where gas or fluid injection leads to cavitation [11, 41], and *indirectly* suggests that several cracks nucleate and grow from the highly deformed cavity wall. This assertion is also supported by the optical images at *t > 360 s* in **FIG4**A where the concavities progressively shrink over time and become invisible.

This time resolved information on cavity size tells us that network scission events must have taken place during the growth stages, but do not give information on the spatial localization of the necessary network scission events that is given by the fluorescence intensity. With the combined real-time information on cavity size and post-mortem fluorescent maps of damage by network scission, we can propose the following scenario:

At the early stages of cavitation, the growth rate is very high (see time-lapsed evolution of the cavity contour in **FIG S4** and steep slope in **FIG2**B), and as a result the elastomer suffers more damage (see high fluorescence near the center of the damaged region in **FIG5**). This observation is consistent with recent investigations on $EA_{0.5}(+)$ fractured under monotonic loading (*i.e.,* mode-I fracture test) where high loading rates and fast crack propagation increase and delocalize damage near the crack surface [35]. As such, it is reasonable to attribute the thick highly fluorescent region (of radius ∼ 50-100 μm and thickness ~25 μm) at the center of a damaged region to the nucleus of a cavity from which multiple crack planes subsequently grow (more slowly) as large and thin petals (thickness ≤10 μm as illustrated by the z-slices in **FIG5**A).

Interestingly, the fluorescent damage regions also exhibit intermediate crack arrest lines (green arrows in **FIG5**C) and out-of-plane crack bifurcations (see **FIG S5**) that suggest discontinuous (*i.e.*, non-steady state) fracture propagation. This behavior is only resolved by visualizing damage with a laser scanning fluorescent microscope in the unloaded configuration and resembles that observed during *stick-slip* fracture in soft materials [42], where unstable fast crack propagation is followed by temporary crack arrest and nucleation of a new crack. It should be noted that cavity growth is driven by the applied energy release rate, which depends on both internal pressure and cavity size and shape [18]. Simulations show that during this stage of stick-slip cavity growth, the pressure inside the cavity progressively decreases [43] but since the energy release rate for crack growth increases with size the cavity can continue to grow despite a decreasing internal pressure, a counterintuitive result. Of course, at some point the pressure becomes too low to further nucleate new side-cracks and the cavity deflates and eventually closes.

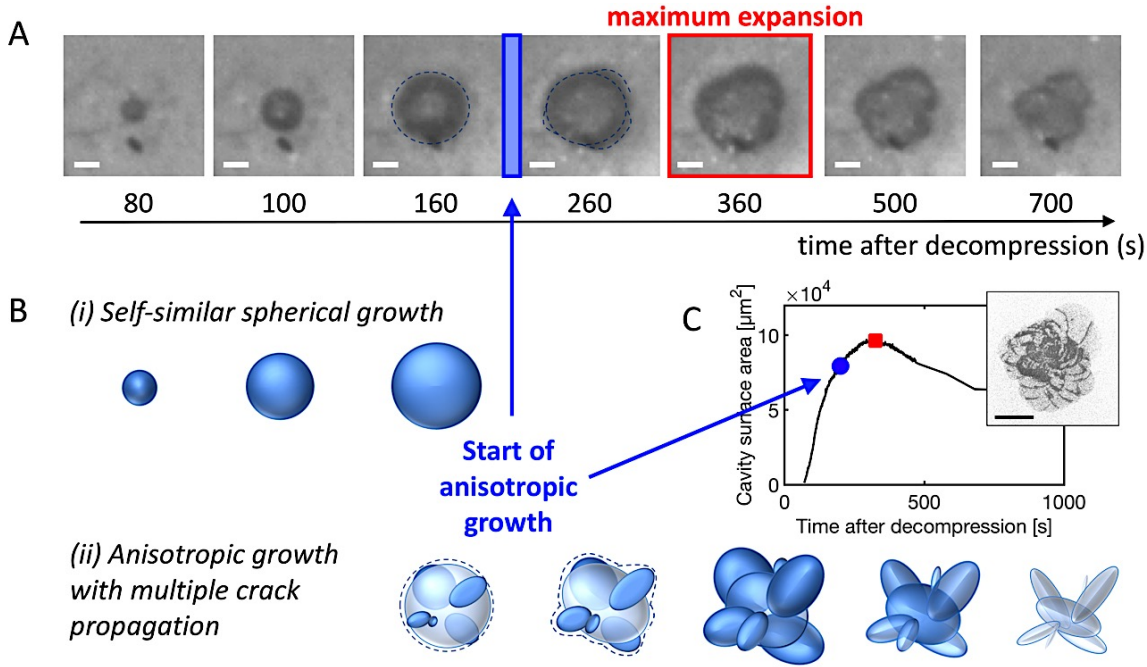

**Figure 4.** (A) *Optical snapshots of cavity n°3 in **FIG3** during growth (white scale bar corresponds to 100 µm). Blue rectangle indicates visual transition from self-similar spherical to anisotropic raspberry-like growth, with dashed blue lines highlighting differences in shape. Red square corresponds to maximum cavity expansion. (B) 3D schematics of suspected cavity growth regimes: (i) self-similar growth of a spherical cavity; (ii) anisotropic raspberry-like growth by propagation of multiple crack planes. (C) Evolution of the cavity 2D projected area. Inlet shows the top view of the associated fluorescent damaged region for comparison with the inflated cavity at 700 s (black scale bar corresponds to 100 µm).*

**Discussion**

Tagging elastomers with mechanofluorescent probes enables non-destructive mapping of network scission in unloaded and pristine looking elastomers after complex 3D loading. This novel information is now discussed and put in perspective with previous work.

First, due to the complex loading configuration involving gas After rapid decompression, the competing rates of solid-gas phase separation and diffusion lead to cavity nucleation and growth under varying pressure and volume [43, 44]. This loading configuration is complex compared to the injection of a gas at constant pressure [11] or volume [39, 41] (*i.e.*, puncture) or the deformation of a confined layer at constant applied displacement [9, 19, 22]. As such, a critical pressure for cavitation is challenging to evaluate. However, $EA_{0.5}(+)$ cavitates in a regime where the saturation pressure $P_{sat}$ exceeds the Young's modulus $E$, as expected from the seminal work of Gent and Tompkins [6, 9]. Unfortunately, the lack of time resolved information given by the mechanofluorescent probes and the ~ 10 µm spatial resolution of the camera makes it challenging to detect unambiguously the very early stages of cavity nucleation. However, the thick regions of high fluorescence in the middle of the damage maps suggest the existence of a 25-50 µm nucleation zone where strain rates were high and damage more isotropic.

In seminal experiments Crosby and co-workers [11, 45] have injected gas into model hydrogels with a needle and demonstrated that a cavity first inflates reversibly and then transitions to irreversible fracture. This transition may appear similar to what we have observed here for rapidly decompressed $EA_{0.5}(+)$, where cavity growth is first self-similar spherical and then anisotropic raspberry-like. However, damage zone mapping reveals that even cavities that apparently only experience self-similar spherical growth over the course of decompression (**FIG 2**, **FIG4**A), do result in damaged zones with planar flower-like morphology (**FIG S6**). This observation is reasonable considering the difference in size between the polymer mesh $O(10nm)$ and a cavity $O(10\text{-}100\mu m)$, and underlines that cavity growth results from irreversible fracture of numerous 2D crack planes.

Kim *et al.* [7] and Poulain *et al.* [22, 46] also have *indirectly* demonstrated that cavitation in PDMS must involve fracture. Poulain *et al.* optically monitored a confined PDMS elastomer layer subjected to quasi-static loading and conducted numerical simulations to demonstrate that the observed conical shape of the formed open cavities is consistent with the growth of penny-shape cracks at *µm* length scales, whereas Kim *et al.* cyclically induced liquid-liquid phase separation in mechanically pre-stretched PDMS gels and monitored the cavities aspect ratio to argue that there must be irreversible damage around the cavity surface. To the best of our knowledge, the non-destructive visualization of damage by network chain scission in rapidly decompressed $EA_{0.5}(+)$ enables *direct* and unprecedented 3D mapping of the precise shape and morphology of the damage zone that results from cavity growth.

Kim *et al.* also argued that the transition from self-similar (with distributed radial damage around the cavity wall) to anisotropic growth (with localized crack propagation) occurs for cavity sizes comparable to the elasto-adhesive length L ~ $G_c/E$ [47]. However, our observations indicate that cavity growth in $EA_{0.5}(+)$ always results from fracture of multiple localized crack planes even for cavities smaller than the elasto-adhesive length of our material, L ~ 180 $\mu m$.

Besides theoretical discrepancies concerning cavity nucleation among different groups, there is a consensus about the existence of different cavity growth regimes and the occurrence of fracture for large cavity size. Interestingly, the direct and non-destructive mapping of damage zones due to cavities in unloaded elastomers suggests a discontinuous crack growth process similar to that of stick-slip crack propagation under monotonic loading. This observation is consistent with a theory recently developed by Cai *et al.* [21] where a spherical cavity with an edge crack can experience both stable and unstable growth depending on the loading conditions and the crack size. This description appears appropriate for the case of rapid decompression of $EA_{0.5}(+)$ elastomers, but it is also useful in unifying other theories of cavitation by fracture [7, 11].

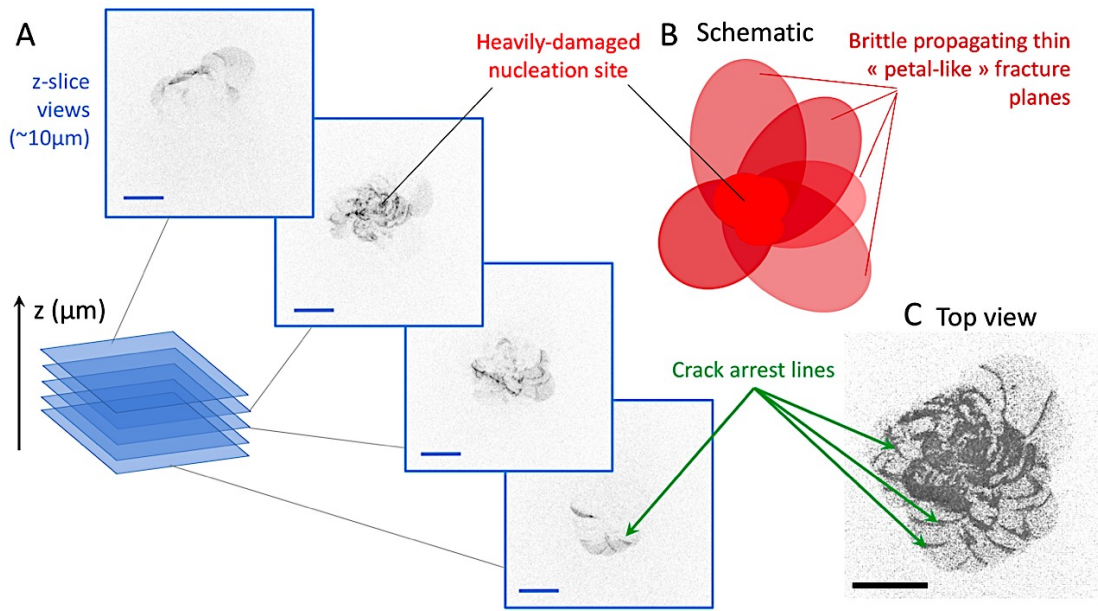

**Figure 5.** (A) *Discretization of the stack of 2D scans that enable the construction of a 3D image with the confocal microscope. Each z-slice scan corresponds to a 10* $\mu$m *thick material volume stressing the thin and planar aspect ratio of the cavity outer crack planes. (B) Schematic representation of the post-mortem top view of a cavity, highlighting a heavily-damaged nucleation site from which multiple fracture planes subsequently grow to form a complex "flower-like" pattern from the superposition of these thin petal-like cracks. (C) Stitched top view of the confocal fluorescent signal mapping of a single representative cavity (cavity n°3 in FIG3). Darker gray corresponds to high fluorescent intensity. Green arrows indicate crack arrest-lines. Blue and black scale bars correspond to 100* $\mu$m.

**Conclusions**

When an elastomer is subjected to a high-pressure gas for a sufficiently long time, gas sorbs until reaching a saturation equilibrium. Upon fast release of that pressure, the gas absorbed within the polymer diffuses out of the material. If diffusion is not fast enough, phase separation occurs leading to the formation of visible pressure-inflated cavities. Even though these cavities eventually disappear as the gas diffuses out, such rapid decompression can lead to macroscopic failure and is of concern for the repeated use of seals in hydrogen storage and transportation. In this work, we tag a poly(ethyl-acrylate) elastomer with a mechano-fluorescent probe that reports irreversible network chain scission by becoming fluorescent through a force activated cycloreversion reaction. Visual monitoring of the cavity inflation/deflation process and 3D post-mortem laser scanning confocal microscopy of the decompressed specimens demonstrate that each cavity results in a localized region of molecular damage in the polymer network.

The fluorescence intensity provides a *direct* proof of the network damage associated with cavity growth and not only enables non-destructive and spatially resolved 3D mapping of internal damage; but also brings unprecedented insights on the shape, deformation and fracture of cavities in elastomers. In our system (and likely in many conventional elastomers), visible cavitation is always accompanied by an irreversible and localized network damage process the size and shape of which correspond to the presence of randomly oriented 2D penny-shape cracks. Additionally, the anisotropic shape observed optically during cavity expansion is closely related to the nucleation and growth of several crack planes originating from a common nucleation site. A careful analysis of the morphology of the damage zone suggests that the process of cavity growth is discontinuous, meaning that it occurs in a stick-slip fashion with stable and unstable stages.

In addition to its usefulness for the fundamental understanding of fracture in elastomers, such a fluorescence-based observation technique has many practical implications. As long as optical detection is possible, the non-destructive visualization of damage by polymer chain scission in elastomers makes it possible to assess the role of multiple cycles of compression/decompression or any other complex loading configuration, on the spatial distribution of internal damage in the polymer network and its impact on seals lifetime.

**Materials and Methods**

**Reagents:** Ethyl acrylate (EA), 1,4-butanediol diacrylate (BDA) and 2-Ethyl acrylate (EA), 1,4-butanediol diacrylate (BDA) and 2-hydroxy-2-methylpropiophenone (HMP) were purchased from Sigma-Aldrich. EA and BDA were purified via elution on activated alumina, and HMP used as received. The mechanofluorescent π-extended anthracene maleimide probe (DACL) was synthesized and purified according to previously established procedures [29]. All reagents were sparged with $N_2$ for 45 *min* and transferred to a $N_2$-filled glovebox.

**Synthesis**: The tagged poly(ethyl acrylate), $EA_{0.5}(+)$ was synthesized according to previously established procedures [33, 35, 48]. In a $N_2$-filled glovebox, EA (4.0 *mL*, 36.68 *mmol*), BDA (33.4 *μL*, 0.18 *mmol*), HMP (66 *μL*, 0.43 *mmol*), and DACL (5.1 *mg*, 0.01 *mmol*) were well-mixed and transferred to a polymerization mold previously sealed with silicone spacers (0.1 *mm*). The polymerization was conducted under UV (10 *μW/cm²*) for 2 h, and the resulting polymer networks dried under vacuum at 50 $^oC$ for 8 *h* to eliminate any residual monomer.

**Material properties**: $EA_{0.5}(+)$ has a glass transition temperature of -18°C as measured by DSC. It exhibits the same mechanical properties with or without mechanophores as previously reported in [35, 36] and illustrated in the stress-stretch curves under uniaxial tension and pure-shear tests in FIG S1 of the SI. The corresponding Young's modulus and mode I fracture toughness were measured to be ~1 MPa and ~180 J/m², respectively.

**Explosive Decompression Tests:** Decompression tests were performed in a pressure chamber (1.77 *L*; diameter 150 *mm*; depth 100 *mm*) engineered to sustain pressures up to 40 *MPa*, and to compress and decompress under controlled ramps. All experiments were conducted at room temperature. The applied pressure cycle can be divided in the following steps: (i) three initial purges of the chamber with gaseous nitrogen up to 1 *MPa* (in order to avoid oxygen/hydrogen mixture), (ii) pressurization at constant rate of 1 *MPa/min* up to a saturation pressure $P_{sat}$ of 5 *MPa*, (iii) dwell at $P_{sat}$ for 10 *h* (in order to bring the sample back to thermal equilibrium and to reach hydrogen's solubility limit in $EA_{0.5}(+)$, according to the Crank's solution for unidirectional gas diffusion through an infinite sheet [49] (iv) decompression to atmospheric pressure $P_{atm}$ at constant rate dP/dt = 10 *MPa/min*, and (v) rest at $P_{atm}$ for 2 *h* in order to track and monitor the delayed evolution of cavitation. To estimate reproducibility, the same test was carried out with two rectangular specimens on two different days, and very similar results were obtained (see **Movies S1** and **S3**).

Note that the diffusion coefficient $D$=1.35e-11 *m²/s* and hydrogen solubility $S$=3.05e-4 *mol $H_2.m^{-3}.MPa^{-1}$*, obtained via permeability measurements, are rather close to that of other elastomers of commercial interest such as EPDM and NBR. These properties confirm that full saturation is achieved prior to decompression, but also validates $EA_{0.5}(+)$ as a good model system for studying cavitation in elastomers under rapid decompression.

Moreover, an additional confirmation that tagging poly(ethyl acrylate) does not impact the mechanical properties as well as the cavitation process is given by the observation of similar cavities (size and growth rate) under identical rapid decompression conditions for specimens with $EA_{0.5}(+)$ and without $EA_{0.5}(-)$ mechanophores (see **Movie S4**).

**Optical Recording**: Pictures of the specimen were taken every second during the first twenty-five minutes after decompression using a wired light and a Sony XCD SX 90CCD camera fitted with an Avenir TV Zoom Lens 12.5-75 *mm* F18. The experimental set-up clearly limited the spatial resolution as the camera had to be located far from the sample (due to the chamber size and front-door thickness – see FIG1B). Moreover, the pixel size had to be compromised to allow for the monitoring of the full specimen surface. Hence, the retained configuration led to an optical resolution ~ 10 *μm*.

The stack of images taken during decompression was processed using the Fiji plug-in of ImageJ® software. First, it was cropped around each selected cavity and a dedicated macro for segmentation was applied to each of them. Particle analysis of the stack of binarized images provided the time evolution of the 2D projected cavity size. Only a selection of cavities was processed, because of limitations resulting from the closeness / overlapping of cavities and from illumination artefacts which may appear at their deformed surface.

**Fluorescent Confocal Microscopy**: All scanning confocal microscope pictures were taken with a Nikon AZ-100/C2+ confocal macroscope. The objective used was an AZ Plan Fluor 5x, with a focal length of 15 *mm*. The objective was not inverted and can zoom from 1x to 8x. Fluorescence was recorded at an image size of 512 x 512 *px* with each pixel containing 12-bit digital units (intensity from 0 to 4096 *a.u.*). The maximal pixel size is 8.15 *μm* and the minimal pixel size is 1.02 *μm*. Confocal mapping reduces out-of-focus light and allows the measurement of intensity in local volumes x×y×z ("voxels") of typically 2.72×2.72×12 *μm³* inside of the material for a 3x zoom setting. A laser with 405 *nm* wavelength was used and the emission signal was recorded from 450 *nm* to 550 *nm*, matching absorption and emission peak of π-extended anthracene moieties (see FIG S7). No specific sample preparation is required except a small dusting-off and careful positioning (facilitated by a biased cut made on one of the specimen edges to keep the same sample orientation than during decompression).

## ASSOCIATED CONTENT

**Supporting Information**

The Supporting Information is available free of charge. It includes a pdf document containing Figures S1 to S7 and legends to Movie S1 to S4, as well as three .avi files corresponding to Movie S1 to S4.

## AUTHOR INFORMATION

**Corresponding Author**

* costantino.creton@espci.psl.eu, morelle.xavier@gmail.com

**Present Addresses**

‡ IMP (UMR 5223 CNRS), INSA Lyon, 69100, Villeurbanne, France.

**Author Contributions**

The project was planned by X.P.M. G.E.S. and C.C. The samples and mechanophores were synthesized by G.E.S. The rapid decompression testing was done by S.C. The confocal optical analysis of the tested specimens was carried out by X.P.M. The visualization data was analyzed by X.P.M. and S.C. and the paper was written by X.P.M., with input from all authors. C.C and S.C. supervised the project. †These authors contributed equally.

## ACKNOWLEDGMENT

This work was supported by the European Research Council (ERC) under the European Union's Horizon 2020 Research and Innovation Program (Grant Agreement N° 695351 – CHEMECH); XPM also acknowledges the financial support of the WBI world excellence post-doctoral fellowship for his two years stay at ESPCI Paris. SC gratefully acknowledges the French Government program "Investissements d'Avenir" LABEX INTERACTIFS (reference ANR-11-LABX-0017-01) which partly funds research conducted in the field of decompression failure at Institut Pprime

SYNOPSIS TOC

Elastomers are classically used as gaskets and seals. If they are saturated with a high-pressure gas and rapidly decompressed, they may suffer from internal damage due to the nucleation and growth of gas bubbles during the decompression. Although insignificant changes in the seals appearance result from the process, internal damage may increase permeability and shorten lifetime, an issue put to the forefront by the needs of safe hydrogen storage and transportation.

We tag elastomers with probes (i.e., mechanophores) that fluoresce upon polymer network failure, and map *the molecular damage resulting from the gas bubbles.* The strong and stable fluorescent signal of the probes gives an unprecedented 3D high resolution view of molecular scale damage compared to other competing non-destructive inspection techniques. We provide direct evidence that, once closed, each bubble results in a disk-like damaged region suggestive of a discontinuous crack growth process caused by the expansion of the gas bubbles. Such details provide unprecedented *molecular rationale for the long-standing consideration of cavitation as a fracture phenomenon* and, more importantly, are a highly promising non-destructive way to characterize *how damage slowly accumulates in soft materials to ultimately compromise functionality and lead to failure*.

*in-situ optical view*

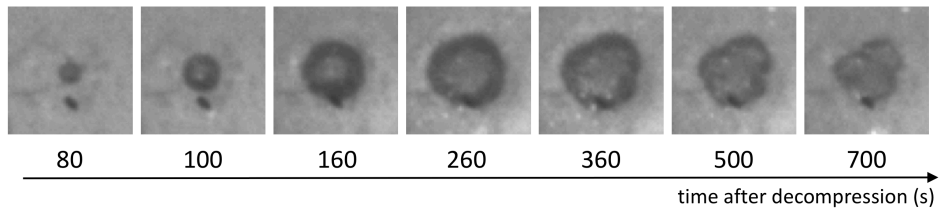

time after decompression (s)

*Image analysis*  *Post-mortem fluorescence*  *Fracture mechanisms*

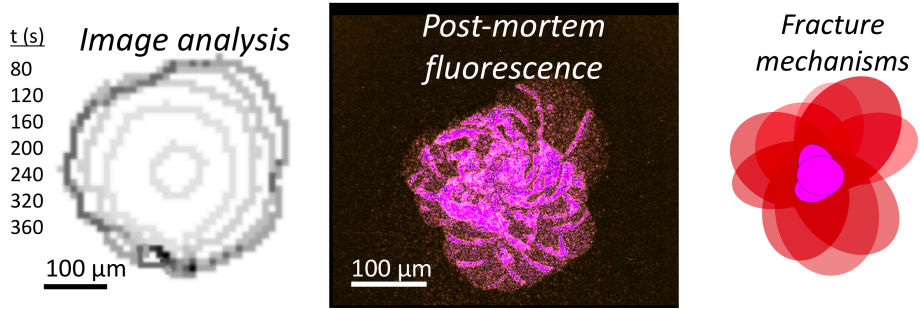

**3D Fluorescent Mapping of Invisible Molecular Damage after Cavitation in Hydrogen Exposed Elastomers**

Xavier P. Morelle*, †,1, Gabriel E. Sanoja†, 1, Sylvie Castagnet[2], and Costantino Creton*,1,3.

[1] SIMM, ESPCI Paris, Université PSL, CNRS, Sorbonne Université, 75005, Paris, France.

[2] Institut Pprime (UPR 3346 CNRS – ENSMA – Université de Poitiers), Department of Physics and Mechanics of Materials, 1 Avenue Clément Ader, BP 40109, 86961 Futuroscope cedex, France

[3] Global Station for Soft Matter, Global Institution for Collaborative Research and Education, Hokkaido University, Sapporo, Japan.

* costantino.creton@espci.psl.eu, morelle.xavier@gmail.com

**This PDF file includes:**

    Figures S1 to S7

    Legends for Movies S1 to S4

**Other supplementary materials for this manuscript include the following:**

    Movies S1 to S4

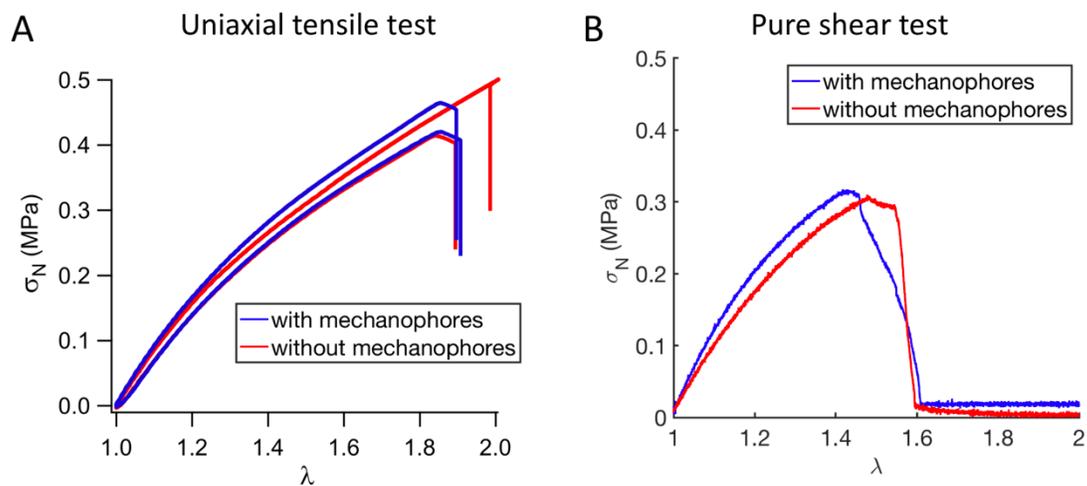

**Fig. S1.** Comparison of the mechanical properties of the ethyl-acrylate polymer network with or without mechanophores. No significant differences in elastic and fracture properties can be noticed both under (A) uniaxial tensile tests; and (B) pure shear tests.

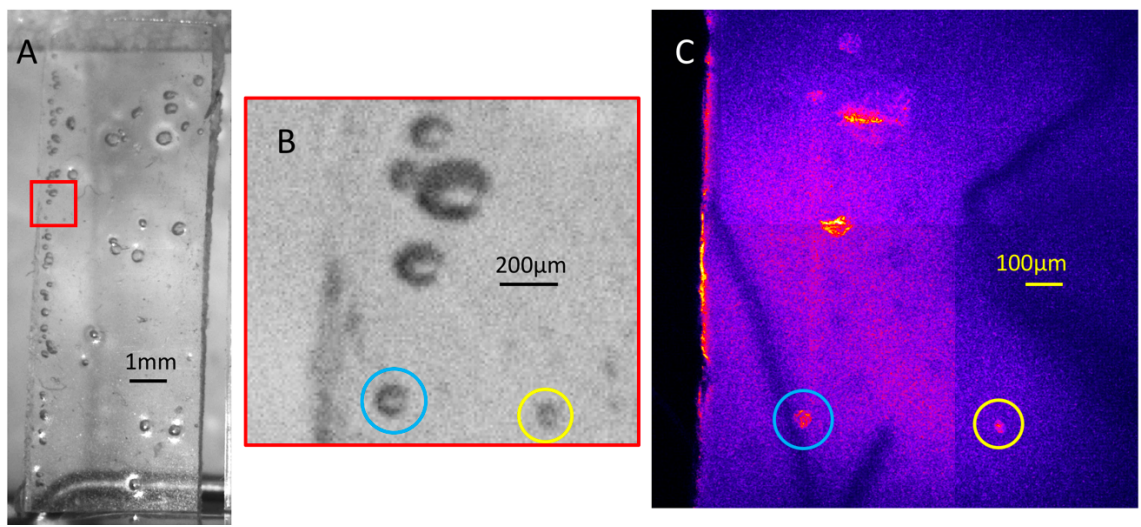

**Fig. S2.** (A) Optical snapshot of second ethyl acrylate specimen (see Movie S2) at maximum cavity expansion during decompression at 10MPa/min after 10h of saturation at a pressure of 5 *MPa* of hydrogen. (B) Zoom on the red square region of (A) where six cavities can be seen, among which the smaller one (yellow circle) is hardly visible by our camera. (C) is the confocal (top view) zoom of the region corresponding to the red square in (A) where a localized fluorescent damage region can be found for each of the six cavities identified in (B). The damaged region of the yellow cavity has a disc shape with a diameter size as small as 30 *μm*.

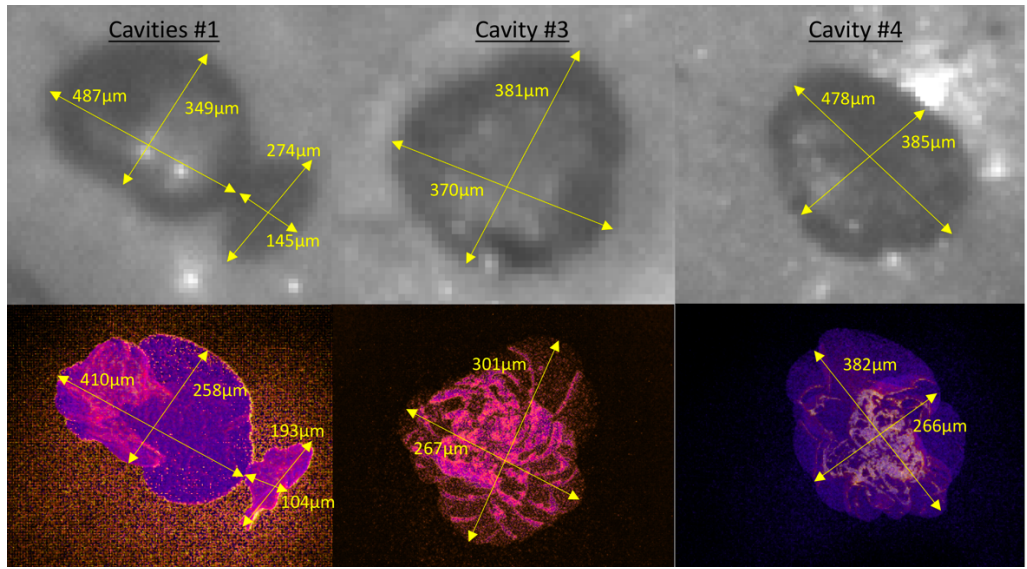

**Fig. S3.** Size comparison between maximum (inflated) cavity through optical visualization (top) and associated damage region, in the deflated state, visualized by fluorescent microscopy. The identified cavities correspond to the number-identification on FIG2A and are fortuitously well-aligned with the specimen xy-plane.

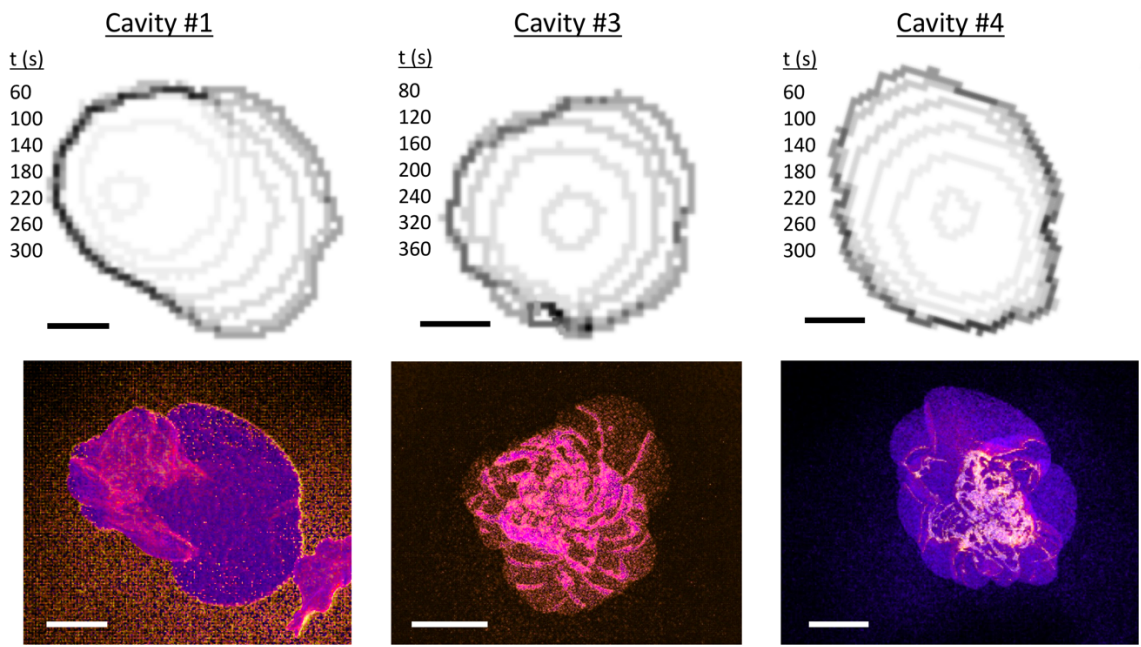

**Fig. S4.** (top) Cavity contour evolution at constant different (constant) time step obtained from optical visualization during decompression (NB: 2D projection limitation thus apply). (bottom) Confocal (artistic) top view of corresponding cavities in order to highlight the difference of fluorescence intensity (and overall contour) between nucleation site and the rest of the flower-like pattern. The 3 visualized cavities correspond to the number-identified on FIG2A, white and black scale bars correspond to 100 $\mu m$.

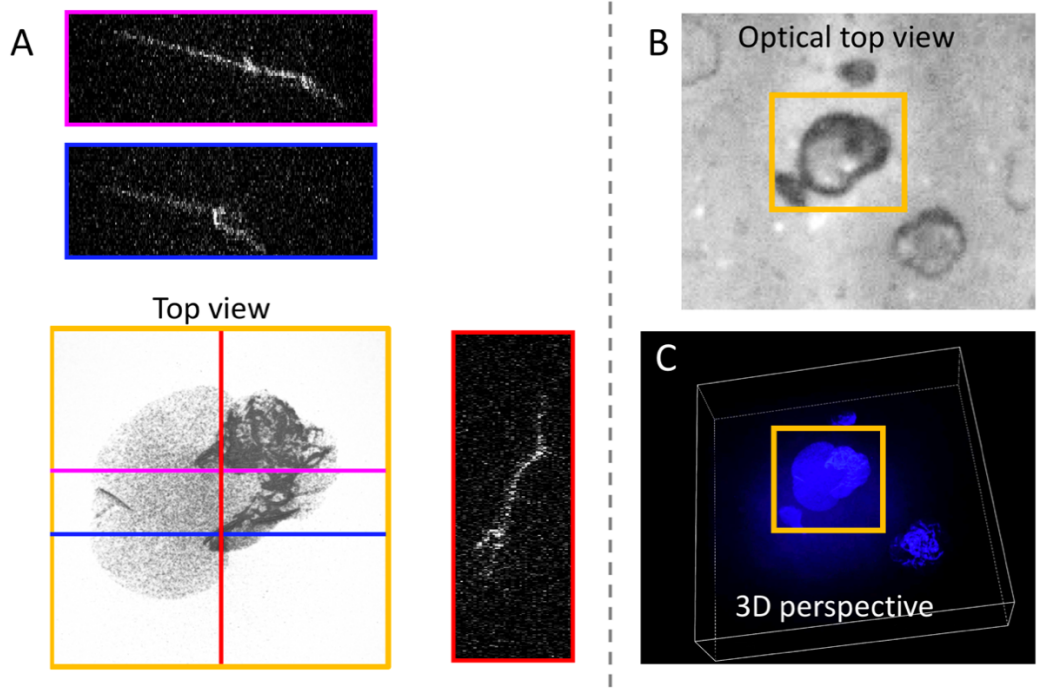

**Fig. S5.** (A) Confocal fluorescent microscopy of the biggest central cavity highlighted with a yellow square in (B) and (C), with corresponding cross-sectional images highlighting local crack bifurcation and tilt of cavity damage region. (B) & (C) are respectively the optical (in the maximum inflated state) and confocal (in the deflated state) view of overall Zone 2 identified in FIG 2.

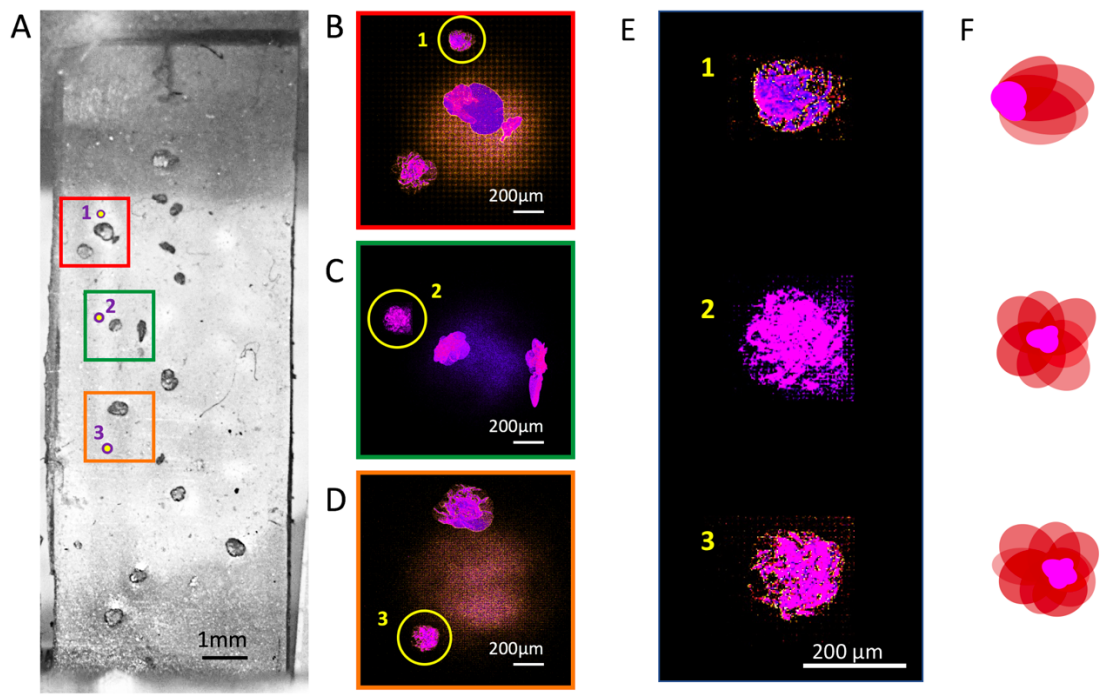

**Fig. S6.** (A) Optical snapshot of our ethyl acrylate elastomer specimen at maximum cavity expansion during decompression, highlighting three colored regions each one with a cavity exhibiting only a self-similar spherical growth (violet circle with yellow dot). (B), (C) and (D) are respectively the confocal (top view) zoom of the corresponding fluorescent damaged regions. The three seemingly spherical cavities are numbered and highlighted with yellow circles with corresponding magnified zoom shown in (E). The fluorescent signal provides a better optical resolution and shows that the damage associated to each of these cavities also corresponds to the superposition of penny-shape cracks forming a flower like pattern (F), hence confirming the same damage growth mechanism than for larger anisotropic cavities. (note that the centered orange or purple halo in (B), (C) and (D) is an optical artifact coming from the fluorescence of the specimen surface.)

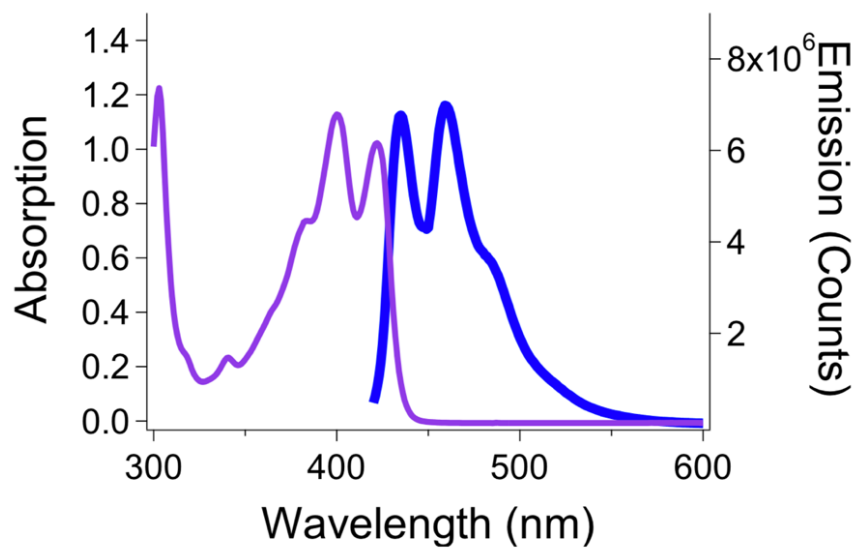

**Fig. S7.** Absorption and emission peak of $\pi$-extended anthracene moieties, data taken from [36]. Photoluminescence and excitation spectra were obtained from an Edinburgh Instrument spectrometer. The absorption spectrum was recorded between 320 *nm* and 450 *nm* for an emission at 460 *nm*. The emission spectrum was recorded between 420 *nm* and 600 *nm*, for an excitation at 405 *nm*. The reference molecule was diluted in ethyl-acetate at $6 \times 10^{-5}$ mol.L$^{-1}$.

**Movie S1 (separate file).** Video made of the snapshots (every 20 s until 400s then every 100s until 1100s) of the different stages during and after a decompression at 10MPa/min of an ethyl-acrylate elastomer saturated at a pressure of 5 MPa of hydrogen for 10 h. The specimen depicted is the same as the one studied through the entire article and totally or partially shown in FIG2, 3 and 4.

**Movie S2 (separate file).** Video made of the 3D mapping with a confocal microscope of the fluorescent damage zones corresponding to the cavities pictured in the red frame (n°1) of Figure 2. The video starts by a progressive z-scan of the region of interest followed by 3D rotation.

**Movie S3 (separate file).** Video made of the snapshots (at various intervals from 18s to 1800s) of the different stages during and after a decompression at 10MPa/min of a second ethyl-acrylate elastomer specimen saturated at a pressure of 5 MPa of hydrogen for 10 h.

**Movie S4 (separate file).** Video made of the snapshots (accelerated 40x) of the different stages during and after a decompression at 5MPa/min of 3 samples of ethyl-acrylate elastomer (the right one with mechanophores and the two left ones without) saturated at a pressure of 5 MPa of hydrogen for 10 h, confirming no significant difference of cavitation resistance (especially no embrittlement) is observed.

**SI References**